\documentclass[
]{ceurart}

\usepackage{listings}

\begin{document}

\copyrightyear{2024}
\copyrightclause{Copyright for this paper by its authors.
  Use permitted under Creative Commons License Attribution 4.0
  International (CC BY 4.0).}


\title{Concept Map Assessment Through Structure Classification}


\author[1]{Laís P. V. Vossen}[%
orcid=0000-0002-1725-0618,
email=lais.vossen@gmail.com,
]
\cormark[1]
\address[1]{Santa Catarina State University, Joinville, Brazil}

\author[1]{Isabela Gasparini}[%
orcid=0000-0002-8094-9261,
email=isagasp@gmail.com,
url=https://www.udesc.br/professor/isabela.gasparini,
]

\author[2]{Elaine H. T. Oliveira}[%
orcid=0000-0003-2884-9359,
email=elaine@icomp.ufam.edu.br
]
\address[2]{Federal University of Amazonas, Manaus, Brazil}

\author[3]{Berrit Czinczel}[%
orcid=0000-0002-4391-7308,
email=czinczel@leibniz-ipn.de
]
\address[3]{IPN| Leibniz Institute for Science and Mathematics Education, Kiel, Germany}

\author[3]{Ute Harms}[%
orcid=0000-0001-6284-9219,
email=harms@leibniz-ipn.de
]

\author[4]{Lukas Menzel}[%
orcid=0000-0002-9870-1109,
email=menzel@studiumdigitale.uni-frankfurt.de
]
\address[4]{Studiumdigitale \& Faculty of Computer Science, Goethe University, Frankfurt am Main, Germany}

\author[5]{Sebastian Gombert}[%
orcid=0000-0001-5598-9547,
email=s.gombert@dipf.de,
url=https://edutec.science/team/sebastian-gombert/
]
\address[5]{DIPF| Leibniz Institute for Research and Information in Education, Frankfurt am Main, Germany}

\author[3]{Knut Neumann}[%
orcid=0000-0002-4391-7308,
email=neumann@leibniz-ipn.de
]

\author[4,5]{Hendrik Drachsler}[%
orcid=0000-0001-8407-5314,
email=h.drachsler@dipf.de,
url=https://www.studiumdigitale.uni-frankfurt.de/91614784/Prof\_\_Dr\_\_Hendrik\_Drachsler\_\_Wissenschaftl\_\_Leitung
]
\cortext[1]{Corresponding author.}

\begin{abstract}
    Due to their versatility, concept maps are used in various educational settings and serve as tools that enable educators to comprehend students' knowledge construction. An essential component for analyzing a concept map is its structure, which can be categorized into three distinct types: spoke, network, and chain. Understanding the predominant structure in a map offers insights into the student's depth of comprehension of the subject. Therefore, this study examined 317 distinct concept map structures, classifying them into one of the three types, and used statistical and descriptive information from the maps to train multiclass classification models. As a result, we achieved an 86\% accuracy in classification using a Decision Tree. This promising outcome can be employed in concept map assessment systems to provide real-time feedback to the student.
\end{abstract}



\conference{EvalLAC 2024}
\maketitle

\section{Introduction}
\label{sec:intro}

Concept Maps (CM) are a robust graphical tool for organizing and representing knowledge, and they can be used for conducting assessments \cite{RuizPrimo2004ExaminingCM}, \cite{StrautmaneMaija:2012}, or even as a study aid that facilitates the observation of knowledge and potential concepts that the student may be struggling to assimilate \cite{review_map:Kirsten2022}, either way, it is considered a robust test of knowledge and understanding of a topic \cite{hay2006}. The assessment methods for CM are diverse, considering both qualitative and quantitative aspects of their construction \cite{Concept_map_assessment:Keppens}. In either form, the analysis of these maps can provide valuable insights into the student's level of learning, highlighting strengths and knowledge gaps. One of the assessment techniques for concept maps, proposed by Kinchin et al. \cite{kinchin2000}, involves categorizing them based on their structure and dividing them into three distinct categories: network, chain, or spoke. These structures proved to be a robust way \cite{KinchinHay2005,KinchinLeij2005} for the educator to visualize the different levels of understanding of a given topic by analyzing the concept map structure that the student created \cite{Gerstner2009,hay2006}. 

Hence, the relevance of concept maps in education is noteworthy, as they serve as a valuable learning instrument that aids educators in the teaching process. Therefore, categorizing the CM by their structure format is crucial to understanding students' rationale, identifying improvement opportunities, and optimizing the learning process. However, the manual categorization process can be time-consuming or even impractical in scenarios with large volumes of produced maps. Therefore, this article addresses the following research question: Can concept maps be classified as Spoke, Chain, or Network based on their features and without relying on human visual interpretation?

To answer this question, we manually classified over 300 concept maps about two distinct topics in one of the three structures. Through feature engineering, we extracted features that characterize each map and used this information to train various machine learning models to find one capable of identifying the maps into either a network, spoke, or chain structure with satisfactory accuracy.

\section{Concept Maps and their Structural types}

\begin{figure} [!h]
\centering
\includegraphics[width=0.4\linewidth]{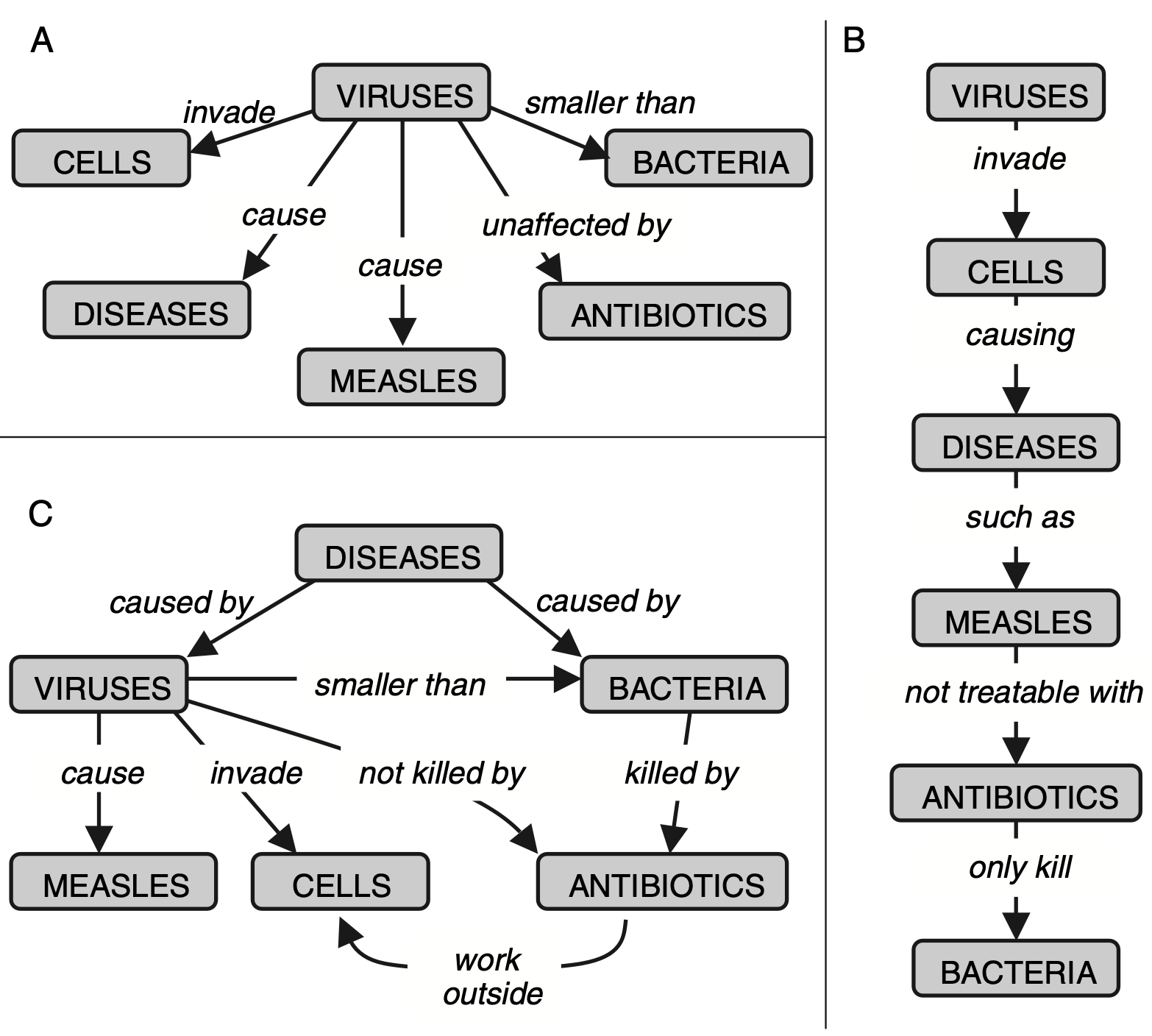}
\caption{Examples of concept maps in three structures: A) spoke; B) chain; C) network. Source: Kinchin et al. (2005) \cite{KinchinHay2005}} 
\label{fig:three_structures}
\end{figure}

The need to establish standards for evaluating these concept maps has arisen. Different evaluation approaches emerged, including quantitative, holistic, similarity-based on a model concept map, or qualitative ones \cite{review_map:Kirsten2022}. Among these evaluation methods, Kinchin et al. \cite{kinchin2000} proposed categorizing concept map structures into three types: spoke, network, and chain, as illustrated in Figure \ref{fig:three_structures}.

An example of a spoke-type map is depicted in Figure \ref{fig:three_structures} in panel A. This type of map exhibits a circular structure in which all the related aspects of the topic are directly linked to the core concept but not directly connected. Such a structure often indicates superficial and underdeveloped knowledge of a subject. Although new knowledge can be more easily integrated, it may only be accessible through reference to the core concept \cite{hay2006,kinchin2000}. In the same figure, panel B illustrates a chain-type map. These structures lack linkage between nodes in different hierarchies and tend to move through available information from start to finish quickly. Consequently, understanding concepts in the middle of the sequence without starting from the beginning may be challenging for students. Additionally, incorporating new knowledge, especially if it disrupts an established sequence, can be difficult \cite{hay2006,kinchin2000}. Finally, panel C exemplifies a network-type map, distinguished by its robust and adaptable cognitive typology. This structure enables the comprehension of a diverse range of concepts through different but related routes of thought, demonstrating students' flexibility in understanding the concept beyond a connection to a core concept or following a specific line of thought \cite{hay2006,kinchin2000}.

\section{The Database}

A collaboration with the Leibniz Institutes DIPF and the IPN provided  data from the use of a custom-made digital concept mapping tool that was used within a ten-week teaching unit \cite{czinczel:species_change}, which was implemented with upper secondary level biology classes in the federal state of Schleswig-Holstein, Germany. Preparation for analysis involved a dataset comprising 317 distinct maps with at least two links between concepts and, therefore, at least three concepts. The maps are divided into two categories: concept maps about the system “school” and concept maps about evolution.

\subsection{Data preparation and cleaning}
\label{subsec:data_cleaning}

The dataset underwent a transformation process to turn the nodes and edges connections into a directed graph structure, allowing the extraction of statistical data about the number of connections which node had. There was also a feature engineering procedure whereby each map was compiled within a single row of the dataset containing the following information:

\begin{itemize}
    \item Map identification: the file name that contained the entire concept map;
    \item Number of nodes: the sum of all nodes in the concept map;
    \item Number of edges: the sum of all edges in the concept map;
    \item Ratio between nodes and edges: calculated by dividing the number of edges by the number of nodes in that map. This variable indicates the overall connectivity of a concept map;
    \item Mean number of edges per node: the mean number of edges that the nodes have, it represents the mean number of connections by node;
    \item Standard deviation of edges per node: the expected variation from the mean number of connections by node;
    \item First, second, and third quartile: the number of connections a node has, below which 25\%, 50\% and 75\% of the smallest number of connections by node lie, respectively;
    \item Maximum number of edges per node: the maximum number of edges connected to a node that happens in the concept map.
\end{itemize}

\subsection{Classification among structures types}

After transforming the data into graph structures, each map was plotted as a directed graph using the igraph library \cite{iGraph}, omitting all labels and thus retaining only the general format of the map, for that reason, the maps' topic didn't influence the classification. Each image was analyzed and classified as a spoke, chain, or network. To accomplish this, we analyzed the most predominant structure of the graph for classification, and at the end of the classification process, 61 CM were classified as chain, 191 as network, and 65 as spoke.


\begin{figure}  [!h]
    \centering
    \includegraphics[width=0.5\linewidth]{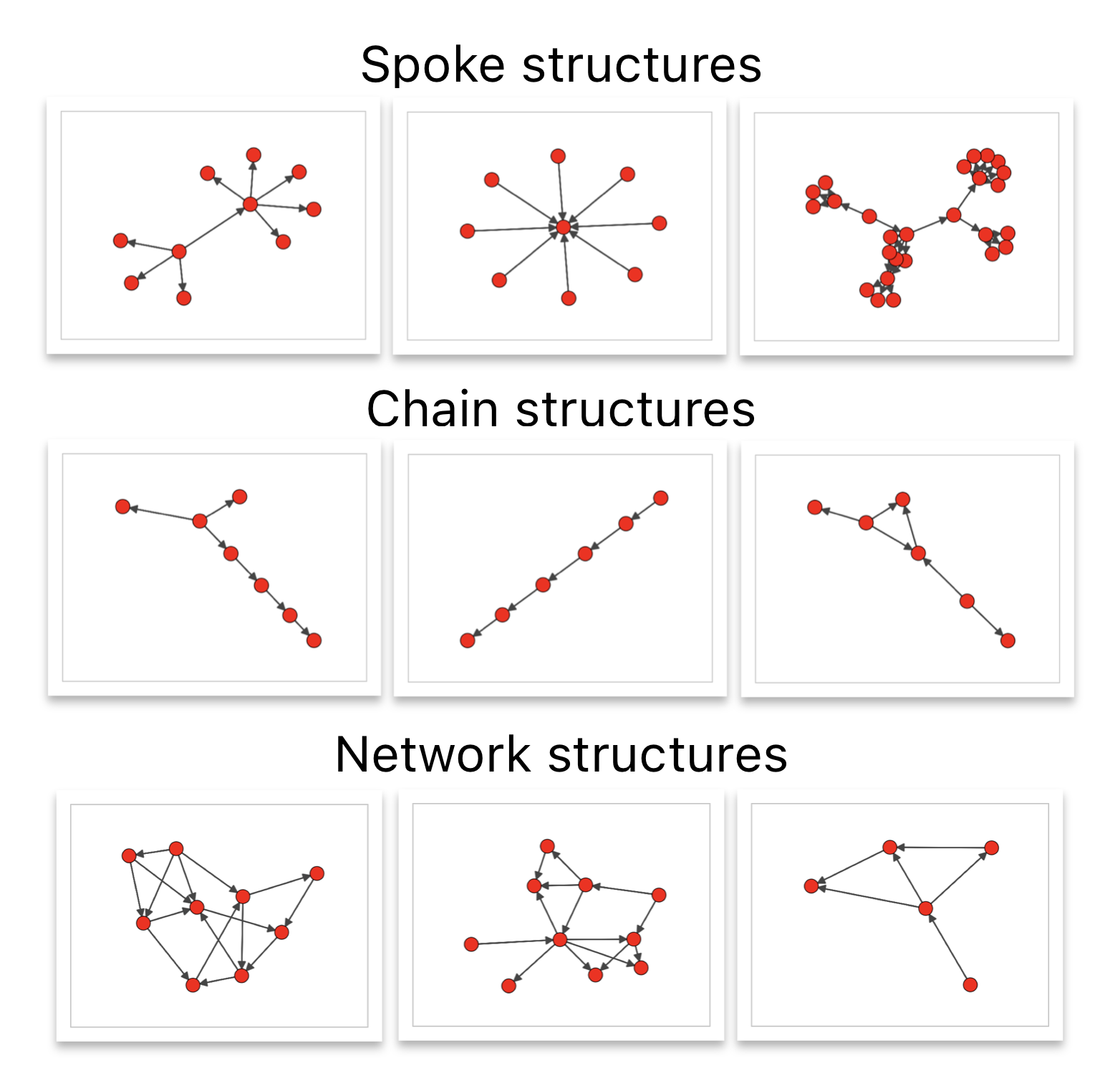}
    \caption{Examples of spoke, chain, and network structures found in the maps}
    \label{fig:structures_example}
\end{figure}

Figure \ref{fig:structures_example} shows some examples of those structures found, in which the first row presents three examples of cases considered as spoke-type structures, with the middle one being the most traditional. The images on the right and left exhibit the main characteristic of spoke structures, featuring key concepts to which various other concepts are connected. However, unlike the traditional image, they have more than one key concept. Similar to the traditional structure, the concepts linked to a key concept lack interconnections among themselves. The second row presents some examples of structures considered chains, with the middle one being the closest to the classic chain structure. The figures on the right and left exhibit the same pattern of having a node connected to the others through a single chain connection. Although these structures deviate from the classic one by having nodes with more ramifications, these nodes are few, and the main characteristic of the lack of connection between concepts at different hierarchical levels is maintained. Finally, the third row presents examples of network structures where highly connected nodes prevail in all cases without a well-defined hierarchical structure. In the dataset, there are several examples of networks with varying degrees of connectivity, and structures that do not exhibit a prevalence of chain or spoke characteristics were considered as networks.

\section{Automatic Concept Map Structure Classification}

To address the research question, the prepared dataset was divided into two parts and balanced to have 100 random samples of each class, with 20\% of the data used as a validation dataset and the remainder used for model training. Subsequently, five multiclass classification models were selected, as presented in Table \ref{tab1}. The models were implemented using the Scikit-learn library \cite{sklearn}, with the default settings for each model. Then, the accuracy was evaluated for each model using cross-validation with the default 5-fold cross-validation method and the F1-Score for each class. Table \ref{tab1} shows that the Decision Tree model outperforms or matches all the metrics, making it the best choice for this particular case. The Permutation Importance method showed that the five most important features were, in descending order of importance: standard deviation of edges per node, mean number of edges per node, third quartile, maximum number of edges per node, and number of edges.


\begin{table}[!h]
\centering
\caption{Model results for classification in chain, spoke or network structure}
\label{tab1}
\begin{tabular}{l|l|l|l|l}
Classifier Model    & Accuracy & F1-Score Chain & F1-Score Network & F1-Score Spoke  \\ 
\hline
Decision Tree       & \textbf{0.86}     & \textbf{0.96}           & \textbf{0.83}           & \textbf{0.93}              \\
Random Forest       & \textbf{0.86}     & 0.94           & 0.79           & \textbf{0.93}             \\
Logistic Regression & 0.78     & 0.86           & 0.69           & 0.77              \\
K Neighbors         & 0.80     & 0.78           & 0.62           & 0.76              \\
Gradient Boosting   & 0.86     & 0.96           & 0.80           & 0.91             
\end{tabular}
\end{table}

\section{Discussion}

This work addressed the research question presented in Section \ref{sec:intro}: \textit{Can concept maps be classified as Spoke, Chain, or Network based on their features and without relying on human visual interpretation?} To achieve this, a thorough cleaning and feature engineering process was employed on the available dataset. As a result, various multiclass classification models were tested, with Decision Tree achieving the highest accuracy of 86\% for classifying a map into the spoke, chain, or network structures.

The results obtained are promising for integrating automated concept map evaluation systems, which could improve educators' routines in utilizing these resources. They could also provide real-time feedback to students regarding the structure of their maps, guiding learning toward further depth. A major challenge of this research lies in the manual classification of maps, as we found no robust guideline to differentiate maps that exhibit characteristics of more than one structure. Therefore, in future work, we suggest a thorough discussion of these classification patterns through detailed analysis of the generated structures, as this may further improve the accuracy and F1-Score results obtained by the models.

\begin{acknowledgments}
 We thank the project ALICE (Analyzing Learning for Individualized Competence development in mathematics and science Education), funded by the Leibniz Society, grant K365/2020. This research is also partially supported by CNPq grant 302959/2023-8 (DT2) and 303443/2023-5 (DT2), and FAPESC Edital nº 48/2022 TO n°2023TR000245.  
\end{acknowledgments}

\bibliography{sample-ceur}


\end{document}